# Uranium oxides investigated by X-ray absorption and emission spectroscopies

M. Magnuson, S. M. Butorin, L. Werme and J. Nordgren
*Department of Physics, Uppsala University, P. O. Box 530, S-751 21 Uppsala, Sweden*

K. Ivanov, J.-H. Guo and D. K. Shuh
*Lawrence Berkeley National Laboratory, Berkeley, California 94720, USA*

## Abstract
X-ray absorption and resonant X-ray emission measurements at the O 1s edge of the uranium oxides $UO_2$, $U_3O_8$ and $UO_3$ are presented. The spectral shapes of the O $K_\alpha$ X-ray emission spectra of $UO_3$ exhibit significant excitation energy dependence, from an asymmetric to a symmetric form, which differs from those of $UO_2$ and $U_3O_8$. This energy dependence is attributed to a significant difference in the oxygen-uranium hybridization between two different sites in the crystal structure of $UO_3$. The spectral shapes of $UO_2$ and $U_3O_8$ are also found to be different but without significant energy dependence. The experimental spectra of the valence and conduction bands of the uranium oxides are compared to the results of electronic structure calculations available in the literature.

## 1. Introduction
Uranium compounds have attracted growing attention from researchers within the last couple of decades. These materials are interesting not only from a technological point of view as the constituents of the nuclear fuel cycle, but also from a scientific point of view since their electronic structures and their macroscopic properties are strongly affected by the partially localized nature of the 5f-electrons. Thus, uranium compounds belong to a class of materials with properties intermediate to those characteristic of localized systems and itinerant systems [1]. Previously, materials quality and the radioactive properties of the actinide compounds have been a limiting factor for more extensive studies of these materials. An additional point of interest particularly concerning the uranium oxides is the connection between their electronic structures, the physical properties of the material and the geometric arrangement of the atoms i.e., the presence of the inequivalent crystallographic sites occupied by chemically identical atoms. Uranium oxides with highly oxidized uranium states contain inequivalent O-U bonds. $UO_2$ (black-brown color), however, has a simple cubic $CaF_2$ crystal structure and is ideally an insulator but is in reality a semiconductor [2]. The increase of the oxygen content and the oxidation states from $UO_2$ to $U_3O_8$ and $UO_3$ results in a decrease of conductivity in the uranium oxides [3]. The formal valency of uranium in $UO_2$ is tetravalent ($U^{4+}$) with $O_h^5$ symmetry around the uranium ion. The lattice structure has identical O sites and has a O-U bond length of ~2.37 Å [4].

In contrast to $UO_2$, the crystal structure is more complicated in the case of the mixed-valency oxide $U_3O_8$ (dark-green color) that contains both $U^{4+}$ and $U^{6+}$ in a 1:2 ratio. In the orthorombic crystal structure of $U_3O_8$, which is an insulator, there are two inequivalent atomic positions for uranium and four different positions for oxygen within the distorted octahedron so that the following uranium-oxygen bonds are formed: $U_I - O_{II} = 2.07$ Å, $U_I - O_{IV} = 2.18$ Å, $U_{II} - O_{III} = 2.07$ Å and $U_{II} - O_I = 2.21$ Å [4]. On the contrary, orthorhombic $\gamma$-$UO_3$, (orange-yellow color) has a formal U valency of $U^{6+}$, and two types of U-O distances are known: short axial oxygen uranyl bonds (~1.79 Å) and longer, planar, equatorial bonds (~2.30 Å) [4,5].

In this paper, we present the results of a resonant X-ray emission investigation at the O 1s edges of a series of uranium oxides; $UO_2$, $U_3O_8$ and $UO_3$. Since the uranium-oxygen systems are relatively complex, the inequivalent site selectivity of the X-ray emission technique makes it particularly useful [6]. We discuss in detail the excitation energy dependence of oxygen $K_\alpha$ emission spectra and the electronic structures of these oxides.

## 2. Experimental details

The $UO_2$, $U_3O_8$ and $UO_3$ samples were powders of 99.8% purity bought from Alpha Aesar (Ward Hill, MA USA). The powders were pressed into indium and attached to an aluminum sample holder. The X-ray absorption (XAS) and X-ray emission measurements at the oxygen 1s edge of these oxides were performed at the undulator beamline 7.0 at the Advanced Light Source in Berkeley, CA. The resolution of the monochromator was about 0.3 eV for the absorption and 0.5 eV for the emission measurements. The XAS spectra were measured at normal incidence in the total electron yield (TEY) mode. X-ray emission spectra were obtained using a fluorescence spectrometer [7] with a resolution of about 0.5 eV. To minimize self-absorption effects in the fluorescence measurements, the incident angle of the photon beam was approximately 20° from the sample surface.

## 3. Results and discussion

Figures 1-3 (upper panels) display the X-ray absorption spectra of $UO_2$, $U_3O_8$ and $UO_3$. The vertical arrows indicate excitation energies at which the resonant X-ray emission spectra were measured. The lower panels show the oxygen $K_\alpha$ emission spectra recorded at various excitation energies throughout the corresponding O 1s-edges. The emission spectra originate from the oxygen 2p -> 1s radiative transitions and reflect the contribution of the occupied oxygen 2p states to the valence band of the uranium oxides. A significant contribution from scattering was not observed in the spectra.

Figure 1 (bottom panel) shows that the oxygen $K_\alpha$ X-ray emission spectra of $UO_2$ have significant asymmetric shapes that do not change much with varying excitation energy. However, a weak shoulder appears at the high-energy side of the spectrum excited with a 561.1 eV incident photon beam. The energy position of the main maximum remains constant for all excitation energies. A similar behavior is also observed for the O $K_\alpha$ emission spectra of $U_3O_8$ in Figure 2. However, in the case of $U_3O_8$, the spectra are broader and the spectral shape is more symmetric than for $UO_2$.

As shown in Figure 3, the situation is different in $UO_3$. In this case, the spectral shapes changes significantly when the excitation energy is changed. As the excitation energy is increased from 530.6 eV, the O $K_\alpha$ spectra transform from an asymmetric to a symmetric shape and broaden. There are two main features to take into account; i) the O $K_\alpha$ spectrum of $UO_3$ excited at 530.6 eV has an asymmetric shape and ii) the $UO_3$ O $K_\alpha$ spectrum excited at 533.7 eV has the most narrow and symmetric line shape.

The results are interpreted by taking into account the structural differences of the studied uranium oxides. It is well known that the unit cell for $UO_2$ contains 12 atoms: 4 crystallographically equivalent U atoms and 8 crystallographycally equivalent O atoms [4,8]. The similarities of the O $K_\alpha$ emission spectra of $UO_2$ for different excitation energies can be explained by the fact that this oxide does not have any inequivalent oxygen sites and as a result, there are no distinct different contributions into the radiative X-ray fluorescence decay processes. The difference of the $K_\alpha$ X-ray emission spectral profiles of $UO_3$ observed at 530.6 eV and 533.7 eV excitation energies can then be attributed to the significant difference in O-U hybridization between the equatorial and axial sites in this oxide. The asymmetric shape in the O $K_\alpha$ spectrum excited at 530.6 eV is a signature of high contribution from the O sites with strong O-U hybridization with the shortest axial (1.79 Å) bonds [6]. On the contrary, the symmetric shape of the O $K_\alpha$ spectrum excited at 533.7 eV is a signature of high contribution of the other oxygen site with weaker hybridization between oxygen and uranium states, forming longer equatorial O-U bonds (2.30 Å) [4,8]. At higher excitation energies, there is a superimposition of contributions from both oxygen sites. The enhancement of the contribution from one of the sites versus the other is likely due to differences in the X-ray absorption cross section at the corresponding energies from the two different sites.

Contrasting to the emission spectra of $UO_2$ and $UO_3$, Figure 2 shows that the O $K_\alpha$ spectra of the more complex $U_3O_8$ system are broader and symmetric and do not exhibit any major changes as the excitation energy is increased despite of the presence of four inequivalent oxygen sites in the crystal structure [4]. An explanation to the fact that there are no distinct features observed in $U_3O_8$ could be the occurrence of both strong mixing of the inequivalent oxygen 2p states between themselves but also the mixing with inequivalent uranium states. This superposition implies that it is impossible to distinctly resolve the participation in the X-ray fluorescence decay process of the occupied states of O and thus the spectra represent the radiative contribution from a mixture of several sites that makes the spectra broader. However, the origin of the small growing shoulder at the high-energy side of the O $K_\alpha$ spectra in Fig. 2 can be attributed to multiple ionization satellites.

Figure 4 shows nonresonant O $K_\alpha$ emission spectra measured at 561.1 eV aligned with O 1s XAS spectra of $UO_2$, $U_3O_8$ and $UO_3$. The emission and absorption spectra were plotted on the common binding-energy scale using the O 1s core-level energy values available in the literature [9,10]. For comparison, calculated O 2p DOS of $UO_2$ using the local spin density approximation (LSDA+U) approach [11] is also shown at the bottom. For a direct comparison to the experiment, the calculated DOS were broadened with a Gaussian of 0.5 eV in order to take into account the instrumental resolution and a Lorentzian of 0.15 eV to take into account the O 1s core-level lifetime width. Some discrepancy between experiment and theory is likely due to electron correlation effects in the U 5f-shell.

By comparing the XAS spectrum of the pure $U^{4+}$ valency oxide ($UO_2$) to the pure $U^{6+}$ valency oxide ($UO_3$) and to the mixed-valency $U^{4+}$ - $U^{6+}$ oxide ($U_3O_8$), it can be observed that they are significantly different. The XAS of $U_3O_8$ cannot be represented as a simple superposition of the XAS spectra of $UO_2$ and $UO_3$. This is clear by taking into account the difference in the absorption cross-sections for the inequivalent oxygen sites. This finding also adds to the knowledge that in the first approximation, the electronic structure of the mixed-valency oxide $U_3O_8$ cannot be described solely on the electronic structures of $UO_2$ and $UO_3$.

As observed in Figure 4, the LSDA+U calculation correctly reproduces the general shape and energy position of the center of gravity for the occupied oxygen 2p states for $UO_2$. However, the shape of the LSDA+U DOS does not match the experimental XAS spectra and the spectral center of gravity of the unoccupied states is shifted towards the Fermi level. This is mainly due to the fact that the calculated band gap (1.1 eV) is highly underestimated in comparison to the experimental one (1.8 eV) [12]. The band-gap in $UO_2$ is of f - f nature and a strong U 5f - O 2p hybridization takes place [13]. This implies that one reason for the discrepancy between the experiment and the calculation is due to an underestimation of the Coulomb site interaction parameter U. Further experimental and theoretical investigations are, therefore, needed in order to clarify this issue.

## 4. Summary

The oxygen sites in the electronic structures of the uranium oxides $UO_2$, $U_3O_8$ and γ-$UO_3$ (6+) have been investigated using oxygen $K_\alpha$ resonant X-ray emission as well as oxygen 1s X-ray absorption spectroscopy. The inequivalent site selectivity of the resonant X-ray emission process is here particularly useful. The excitation energy dependence observed in the oxygen $K_\alpha$ emission spectra for $UO_3$ is attributed to the presence of significant differences in oxygen-uranium hybridization between two different sites. The absence of an energy dependence in the spectral shape of the oxygen $K_\alpha$ X-ray emission spectra of the mixed-valency oxide $U_3O_8$ is explained by the strong mixture and superposition of inequivalent oxygen 2p states that prevents the separation of their distinct contributions of the radiative oxygen 2p -> oxygen 1s transitions. This implies that complementary experimental methods are needed for a deeper understanding of the electronic structures in these rather complex uranium oxide systems.

## Acknowledgments

This work was supported by the Swedish Natural Science Research Council and the Göran Gustafsson Foundation for Research in Natural Science and Medicine. This work was performed in part at the ALS, which is operated by the U.S. Department of Energy (DOE), Office of Basic Energy Sciences (BES), Division of Materials Sciences under Contract No. DE-AC03-76SF00098 at LBNL. This work was also supported by the DOE-BES Chemical Sciences Division at LBNL under the same contract.

## Figure captions

**Figure 1:** O 1s XAS spectrum of $UO_2$ at the O 1s edge (top panel). The arrows in the XAS spectrum indicate the excitation energies used in the X-ray emission measurements. The bottom panel shows resonant O $K_{\alpha}$ emission spectra of $UO_2$ at a number of excitation energies across the O 1s edge.

**Figure 2:** O 1s XAS spectrum of $U_3O_8$ at the O 1s edge (top panel). The arrows in the XAS spectrum indicate the excitation energies used in the X-ray emission measurements. The bottom panel shows resonant O $K_{\alpha}$ emission spectra of $U_3O_8$ at a number of excitation energies across the O 1s edge.

**Figure 3:** O 1s XAS spectrum of $UO_3$ at the O 1s edge (top panel). The arrows in the XAS spectrum indicate the excitation energies used in the X-ray emission measurements. The bottom panel shows resonant O $K_a$ emission spectra of $UO_3$ at a number of excitation energies across the O 1s edge.

**Figure 4:** O $K_{\alpha}$ X-ray emission and O 1s XAS spectra of $UO_3$, $U_3O_8$ and $UO_2$ on a binding energy scale. All spectra were measured nonresonantly at 561.1 eV excitation energy. The XPS binding energies used are 530.1 eV for $UO_3$, 530.2 eV for $U_3O_8$ and 529.7 eV for $UO_2$ [9,10].

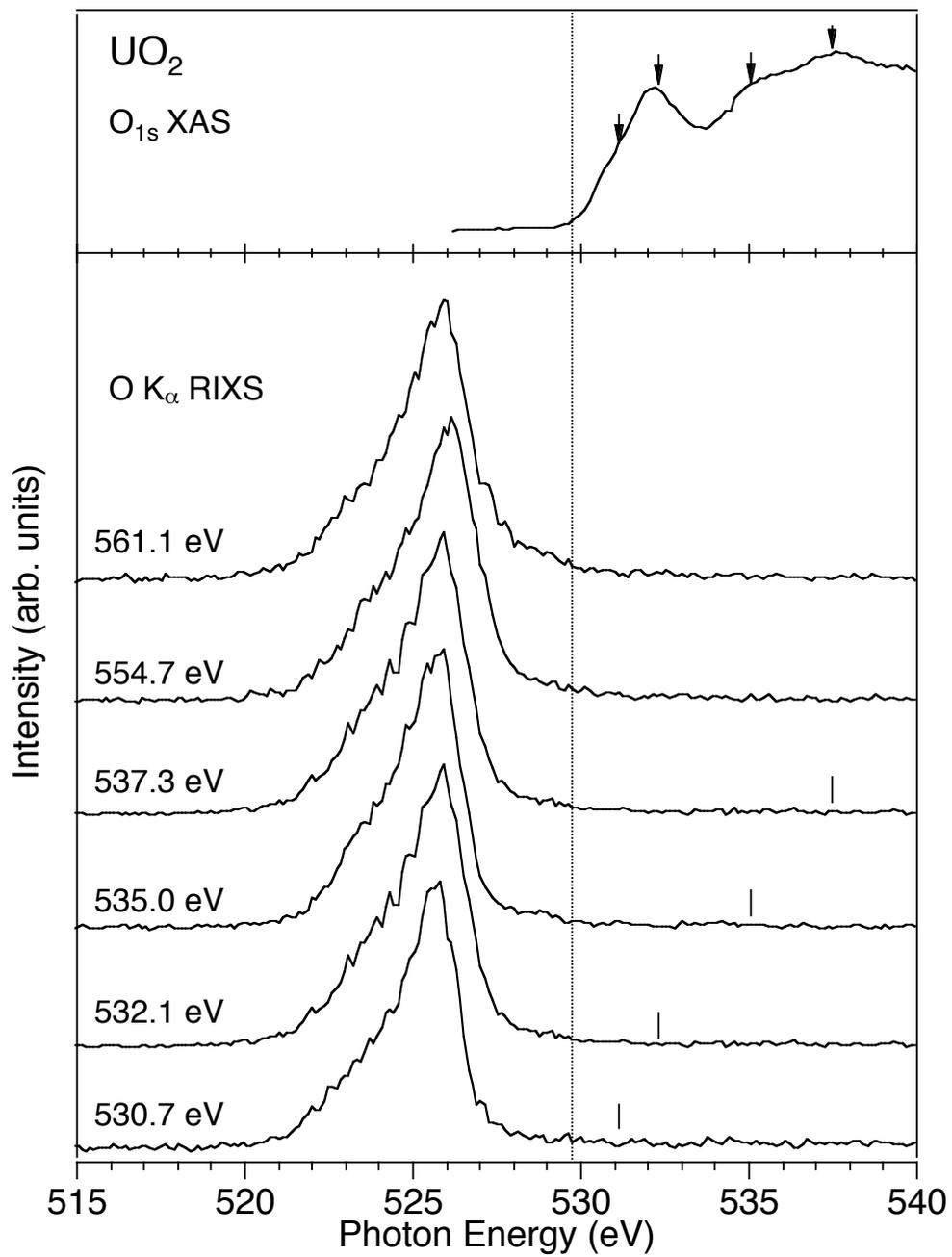

**Fig. 1**

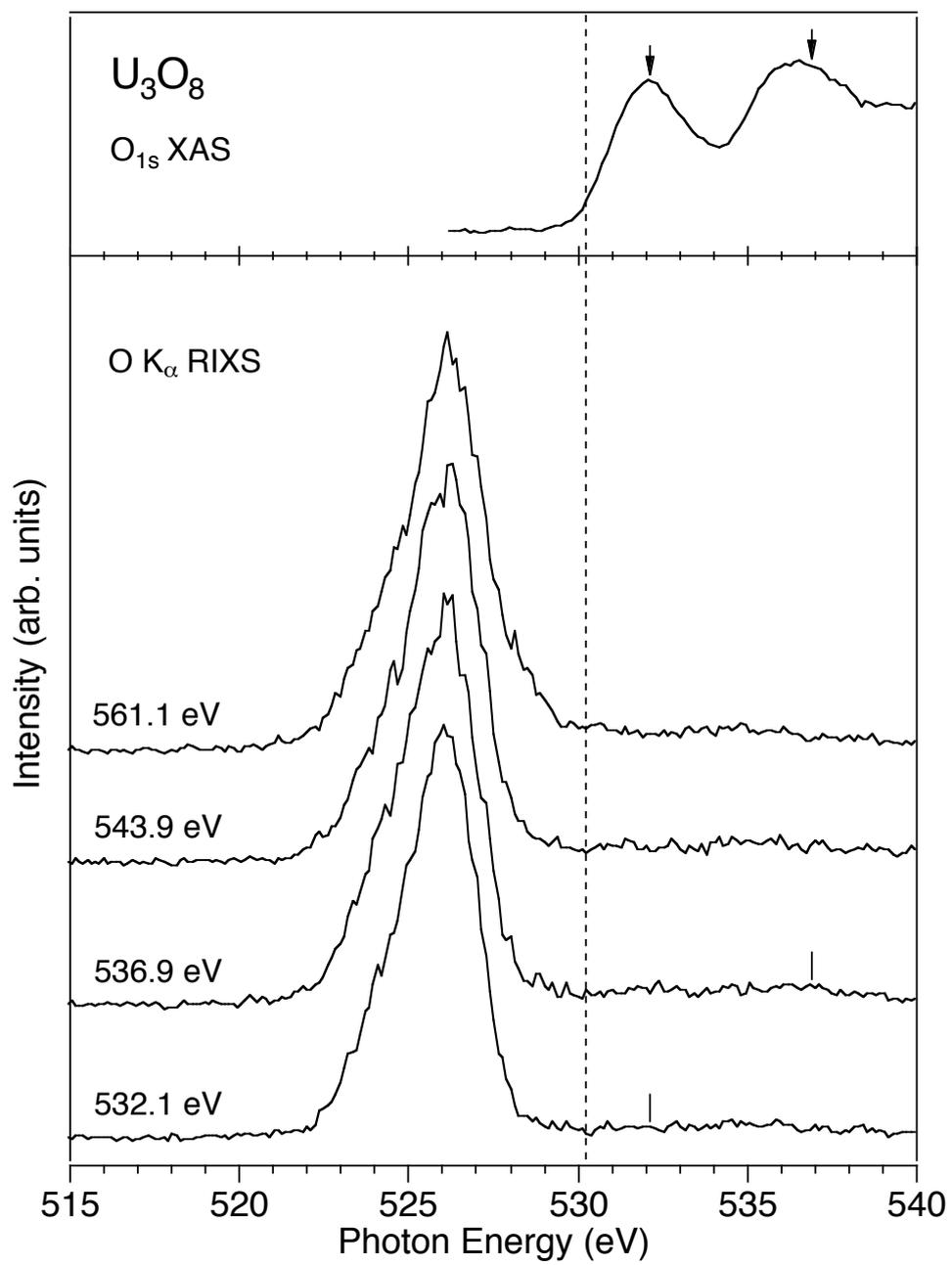

**Fig. 2**

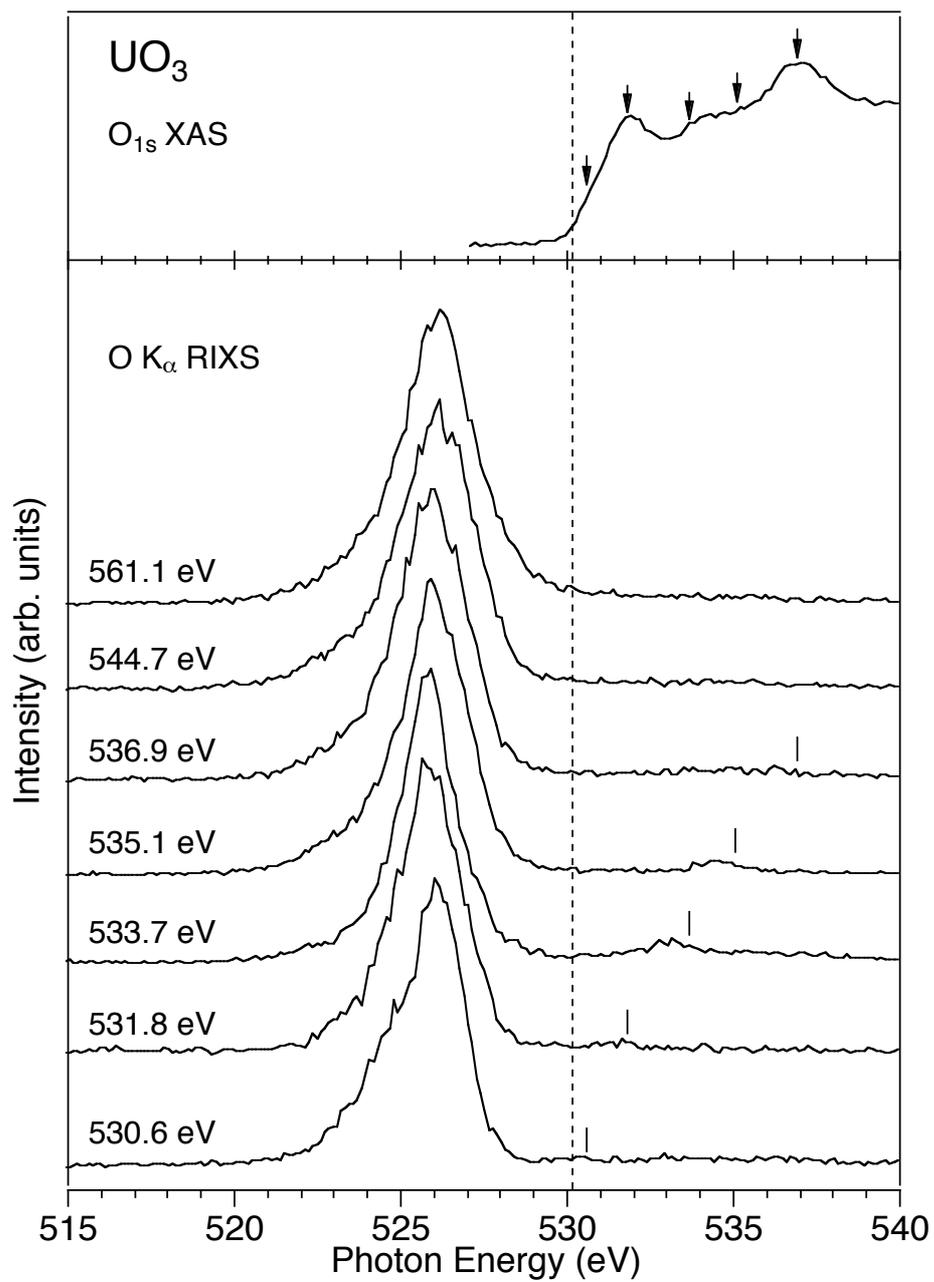

**UO₃**

O₁ₛ XAS

O Kα RIXS

561.1 eV

544.7 eV

536.9 eV

535.1 eV

533.7 eV

531.8 eV

530.6 eV

**Fig. 3**

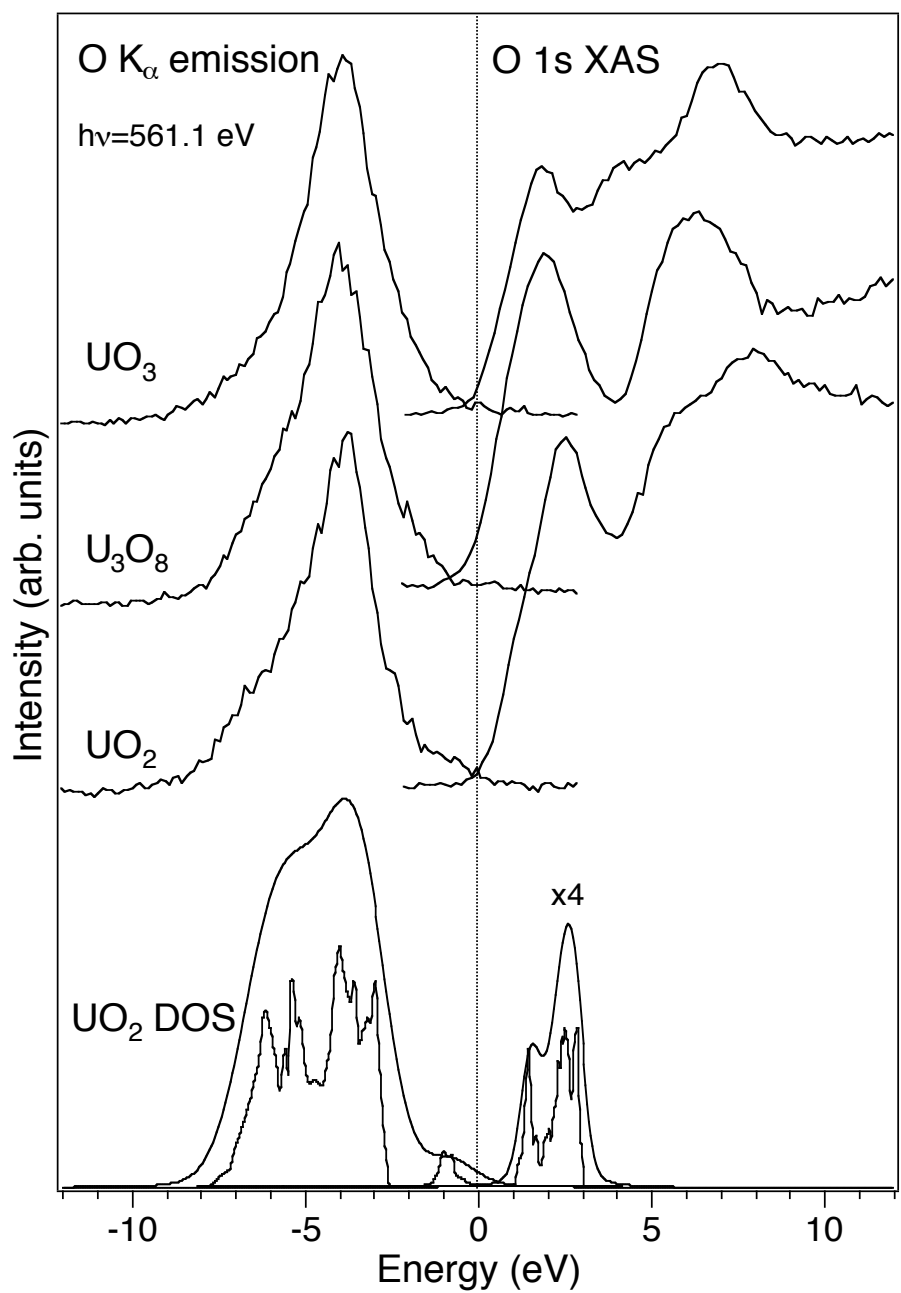

**Fig. 4**